\pdfoutput=1
\documentclass[11pt]{article}  
\usepackage{times}
\usepackage{graphicx}

\usepackage{color}

\usepackage[small]{caption}
\usepackage[latin1]{inputenc}
\usepackage[english]{babel}
\usepackage{amsmath,amsfonts,fancyhdr}
\usepackage{amssymb}	
\usepackage{fancyhdr}
\usepackage{booktabs}
\usepackage{epsfig}
\usepackage{subfigure}
\usepackage{graphicx}
\usepackage[sort&compress,numbers]{natbib}
\usepackage{color}

\newcommand{\be}{\begin{equation}}
\newcommand{\ee}{\end{equation}}

\newcommand{\gv}[1]{\ensuremath{\mbox{\boldmath$ #1 $}}} 
\newcommand{\avg}[1]{\left< #1 \right>} 
\newcommand{\pd}[2]{\frac{\partial #1}{\partial #2}} 
\newcommand{\pdd}[2]{\frac{\partial^2 #1}{\partial #2^2}} 
\newcommand{\laplacian}[1]{\gv{\nabla}^2 #1} 
\let\baraccent=\= 
\renewcommand{\=}[1]{\stackrel{#1}{=}} 

\newcommand{\eref}[1]{Eq.~(\ref{#1})}
\newcommand{\fref}[1]{Fig.~(\ref{#1})}
\begin{document}
\begin{centering}
\noindent 

\long\def\symbolfootnote[#1]#2{\begingroup%
\def\thefootnote{\fnsymbol{footnote}}\footnote[#1]{#2}\endgroup}

{\LARGE \textbf{Low Frequency Sound Propagation in Lipid \\Membranes\\*[0.2cm]}}
\noindent {\large Lars D. Mosgaard, Andrew D. Jackson and Thomas Heimburg\symbolfootnote[1]{corresponding author. Email address: theimbu@nbi.dk}}\\*[0.2cm]
\noindent {\large The Niels Bohr Institute, University of Copenhagen, Copenhagen, Denmark}\\*[0.2cm]
\end{centering}

\rule{0.9\textwidth}{0.4pt}
\begin{abstract}
\noindent In the recent years we have shown that cylindrical biological membranes such as nerve axons under physiological conditions are able to support stable electromechanical pulses called solitons. These pulses share many similarities with the nervous impulse, e.g., the propagation velocity as well as the measured reversible heat production and changes in thickness and length that cannot be explained with traditional nerve models. A necessary condition for solitary pulse propagation is the simultaneous existence of nonlinearity and dispersion, i.e., the dependence of the speed of sound on density and frequency. A prerequisite for the nonlinearity is the presence of a chain melting transition close to physiological temperatures. The transition causes a density dependence of the elastic constants which can easily be determined by experiment. The frequency dependence is more difficult to determine. The typical time scale of a nerve pulse is 1 ms, corresponding to a characteristic frequency in the range up to one kHz.  Dispersion in the sub-kHz regime is difficult to measure due to the very long wave lengths involved. In this contribution we address theoretically the dispersion of the speed of sound in lipid membranes and relate it to experimentally accessible relaxation times by using linear response theory. This ultimately leads to an extension of the differential equation for soliton propagation.\\*[0.2cm]
\end{abstract}


\noindent \textit {Abbreviations:} DSC - differential scanning calorimetry ; DPPC - dipalmitoyl phosphatidylcholine

\
\tableofcontents

\rule{0.9\textwidth}{0.4pt}


\section{Introduction}
Biological membranes are ubiquitous in the living world. Despite their diversity in composition, membranes of different cells or organelles are remarkably similar in structure and exhibit similar thermodynamic properties. They exist as thin, almost two-dimensional lipid bilayers whose primary function is to separate the interior of cells and organelles (sub-cellular compartment) from their external environments.  This separation leads in turn to the creation of chemical and biological gradients which play a pivotal role in many cellular and sub-cellular processes, e.g. Adenosine Tri-Phosphate (ATP) production. A particularly important feature of biomembranes is  the propagation of voltage signals in the axons of neurons which allows cells to communicate quickly over long distances, an ability that is vital for higher lifeforms such as animals \cite{Hodgkin1952,Heimburg2005c}. 
\\\\
Biological membranes exhibit a phase-transition between an ordered and a disordered lipid phase near physiological conditions \cite{Melchior1970}. It has been shown that organisms alter their detailed lipid composition in order to maintain the temperature of this phase-transition despite different growth conditions \cite{Heimburg2007a,Hazel1979,DeLong1985}. The biological implications of membrane phase-transitions continue to be an area of active research. Near a phase transition the behavior of the membrane changes quite drastically: The thermodynamic susceptibilities, such as heat capacity and compressibility, display a maximum and the characteristic relaxation times of the membrane show a drastic slowing down \cite{Tsong1976, Tsong1977, Mitaku1982, vanOsdol1989, Grabitz2002}. 
\\\\
The melting transition in lipid membrane is accompanied by a significant change of the lateral density by about -20\%. Thus, the elastic constants are not only temperature dependent but they are also sensitive functions of density. Together with the observed frequency dependence of the elastic constants (dispersion) this leads to the possibility of localized solitary pulse (or soliton) propagation in biomembrane cylinders such as nerve axons. With the emergence of the Soliton theory for nerve pulse propagation, the investigation of sound propagation in lipid membranes close to the lipid melting transition has become an important issue \cite{Heimburg2005c}. The Soliton model describes nerve signals as the propagation of adiabatic localized density pulses in the nerve axon membrane. This view is based on macroscopic thermodynamics arguments in contrast to the well-known Hodgkin-Huxley model for the action potential that is based on the non-adiabatic electrical properties of single protein molecules (ion channels). Using this alternative model, we have been able to make correct predictions regarding the propagation velocity of the nerve signal in myelinated nerves, along with a number of new predictions regarding the excitation of nerves and the role of general anesthetics \cite{Heimburg2007b}. In addition, the Soliton model explains a number of observations about nerve signal propagation which are not included in the Hodgkin and Huxley model, such as changes in the thickness of the membrane, changes in the length of the nerve and the existence of phase transition phenomena \cite{Andersen2009}. The solitary wave is a sound phenomenon which can take place in media displaying dispersion and non-linearity in the density. Both of these criteria are met close to the main lipid transition. However, the magnitude of dispersion in the frequency regime of interest for nerve pulses (up to 1 $kHz$) is unknown \cite{Heimburg2005c}. Exploring sound propagation in lipid membranes is thus an important task for improving our understanding of mechanical pulse propagation in nerves.   All previous attempts to explore sound propagation in lipid membranes have focused on the ultrasonic regime \cite{Mitaku1982,Halstenberg1998,Schrader2002,Halstenberg2003}, and it has clearly been demonstrated the dispersion exists in this frequency regime. Furthermore, the low frequency limit of the adiabatic compressibility of membranes (which determines the sound velocity) is equal to the isothermal compressibility, which is significantly larger than the compressibility in the Mhz regime. With the additional knowledge that relaxation times in biomembranes are of the order of milliseconds to se\-conds, it is quite plausible to expect significant dispersion effects in the frequent regime up to 1 kHz.
 
Theoretical efforts to describe sound propagation in lipid membranes near the lipid melting transition in the ultra-sonic regime have been based on scaling theory, which assumes critical relaxation behavior during the transition \cite{Bhattacharjee1997, Halstenberg2003}. However, a number low frequency experiments, pressure jump experiments \cite{Grabitz2002,Seeger2007} and stationary perturbation techniques \cite{vanOsdol1989, vanOsdol1991a} all show non-critical relaxation dynamics. These findings have led us to propose a non-critical thermodynamical description of sound propagation in lipid membranes near the lipid melting transition for low frequencies based on linear response theory. 

In this article we will present a theoretical derivation of the magnitude of dispersion for membranes close to lipid melting transitions. The goal is to modify the wave equation for solitons in biomembranes. This will ultimately lead to a natural time scale for the pulse length, which we will explore in future work.


\section{The propagating soliton in nerve membranes}
\begin{figure}[!h]
\centering
\includegraphics[width= 0.5 \linewidth]{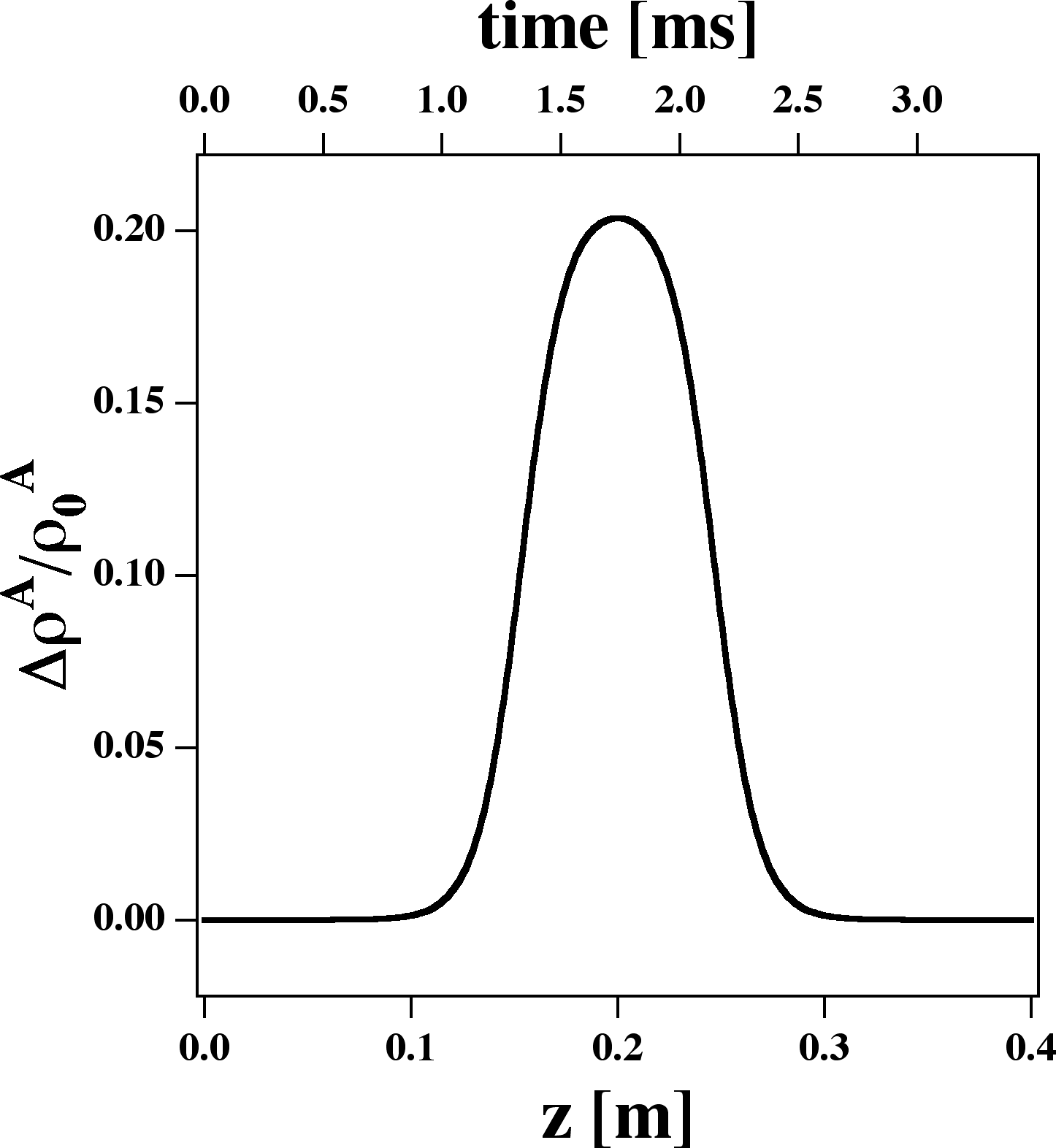}
\caption{\small \textbf{The propagating soliton using parameters appropriate for unilamellar DPPC vesicles and a dispersion constant $h=2$ m$^4$/s$^2$ (from \cite{Heimburg2008}). The soliton has a width of about 10 cm and a duration of about 1\,ms, which is very similar to action potentials in myelinated nerves.}}
\label{soliton}
\end{figure}

In the following we present the hydrodynamic equations that govern the propagation of density waves in cylindrical membranes in general, and in nerve membranes close to the chain melting transition in particular. \\
In it simplest formulation the wave equation for compressible fluids assumes the form\footnote{A derivation of the equation of sound, based on fluid dynamics, can be found in \cite{Landau1987}. There are two basic assumptions in the derivation of the equation of sound: Perturbations are small, and sound propagation is an adiabatic process.}:
\be
\pdd{\rho}{t}=\nabla (c^2 \nabla \rho)
\label{sound0}
\ee
where
\be
c=\sqrt{\left(\pd{p}{\rho}\right)}_{S,0}=\frac{1}{\sqrt{\kappa_S \rho}}
\label{speed0}
\ee
is the speed of sound for low amplitude waves ($\Delta \rho \ll \rho_0$), $\kappa_S$ is the adiabatic compressibility, and $\rho(x,t)$ is the density.   If the speed of sound is roughly independent of density, this equation simplifies to 
\be
\pdd{\rho}{t}=c^2 \, \nabla^2 \rho\,.
\label{sound2}
\ee
The wave equation in one dimension is then given by
\be
\pdd{\rho}{t}=\frac{\partial}{\partial x}\left(c^2\frac{\partial}{\partial x} \rho\right)\,.
\label{sound3}
\ee
For low amplitude sound we further assume that there is dispersion of the form
\be
c^2=c_0^2+h_0\omega^2 + ...\,,
\label{dispersion0}
\ee
which corresponds to a Taylor expansion of the sound velocity with respect to frequency. The parameter $h_0$ indicates the magnitude of the dispersion. Due to symmetry arguments, only even power terms appear in this expansion.
One way to generate this frequency dependence is to add a dispersion term to the wave equation 
\be
\pdd{\rho}{t}=\frac{\partial}{\partial x}\left(c^2\frac{\partial}{\partial x} \rho\right)-h\frac{\partial^4}{\partial x^4}\rho \, .
\label{sound4}
\ee
The density of a small amplitude plane wave can be written as  
\be
\rho(x, t)= \rho_0 + \Delta \rho \ \ \ {\rm with}\ \ \ \Delta \rho = A\sin(kx-\omega t)\equiv A\sin(k(x-ct)) \, .
\label{wave0}
\ee
The amplitude of this plane wave is $A$, and its velocity is with $c=\omega/k$. Inserting this into Eq. \ref{sound3} yields the dispersion relation in Eq. (\ref{dispersion0}) with $h_0=h/c_0^2$.  We have shown experimentally that the sound velocity close to melting transitions in lipid membranes is a sensitive nonlinear function of density. Thus, we expand 
\be
c^2=c_0^2+p\Delta \rho+q(\Delta \rho)^2 + ... \, .
\ee
The parameters $p$ and $q$ describe the nonlinear elastic properties of membranes.  At temperatures slightly above the melting transition, lipid membranes have negative values for the parameter $p$ and positive values for the parameter $q$.  The final wave equation is given by
\be
\pdd{\rho}{t}=\frac{\partial}{\partial x}\left((c_0^2+p\Delta \rho+q(\Delta \rho)^2) \frac{\partial}{\partial x} \rho\right)-h\frac{\partial^4}{\partial x^4}\rho \, .
\label{sound5}
\ee
We have shown that this equation possesses analytical solitary solutions that in many aspects resemble the nerve pulse (see Fig. \ref{soliton}).\\

While the above equation makes use of the fact that the speed of sound is a known function of density, the dispersion constant $h$ must be regarded as an adjustable parameter due to the absence of quantitative empirical data regarding dispersion in the low frequency regime. The magnitude of $h$ sets the width and the time scale of the mechanical pulse. In previous publications it was adjusted to $h=2 m^4/s^2$ in order to match the observed width of the nerve pulse, which is about 10 cm. However, we will argue below that $h$ is expected to be density dependent and that its functional form can be approximated using experimental knowledge about relaxation time scales and elastic constants. This will ultimately lead to a wave equation for the mechanical pulse in nerve axons that is free of adjustable parameters and has a time scale that is 
fixed by the thermodynamics of the system.

\section{Brief Overview of Sound}
Sound is a propagating low amplitude density wave in compressible medium which, due to its adiabatic nature, is accompanied by a corresponding temperature wave. The equation governing sound propagation is universal. This generality implies that sound propagation is determined solely by the macroscopic thermodynamical properties of the system. 

As mentioned above, the equation of sound for low amplitude waves has the form:
\be
\pdd{\rho}{t}=c^2 \laplacian \rho \, . \nonumber
\ee
The general solution has the form:
\be
\rho=A\exp(i \omega(t-x/\hat{c}))\,,
\label{sound_solution}
\ee
which is merely eq. (\ref{wave0}) in complex notation.  Due to dispersion and the absorption of sound in a real medium, the effective speed of sound, $\hat{c}$, is a complex quantity. The real part of the speed of sound will have a phase shift (as a result of dispersion), the imaginary part will lead to a decrease in the amplitude or intensity of the sound as it propagates (attenuation). This can be seen by inserting the complex speed of sound into \eref{sound_solution}.
\be
\rho=A \exp \left( i\omega \left( t-x Re (\hat{c})/|\hat{c}|^2 \right)\right) \, \exp \left( -x \omega Im(\hat{c})/|\hat{c}|^2 \right) \, ,
\ee
where
\be
u = \left( \frac{Re(\hat{c})}{| \hat{c}|^2}\right)^{-1}
\label{real-speed}
\ee
is the effective speed of sound which would be measured in experiment. 
\\\\
In 1928, Herzfeld and Rice extended the theory of sound by arguing that internal vibrational modes of polyatomic molecules require time to approach thermal equilibrium with translational degrees of freedom \cite{Herzfeld1928}. If the time scale of the density (or pressure) perturbation is similar to or less than the time scale of these internal relaxation times, the temperature response of the system will lag behind that of the perturbation. This will prevent the internal degrees freedom from taking up all the heat and will result in a decrease in the effective heat capacity.\footnote{Note that the effective heat capacity will be referred to as the dynamic heat capacity.} This decrease in the effective heat capacity results in hysteresis and in dissipation of heat. \\
In 1962, Fixman applied the basic ideas of Herzfeld and Rice to describe the viscosity of critical mixtures \cite{Fixman1962}. He was motivated by the intimate relation between viscosity and attenuation. Critical mixtures of fluids display a second-order transition which is indicated by a critical slow-down of the relaxation rates of the order-parameters. In contrast to Herzfeld and Rice, Fixman did not limit his attention to the rates of translational and internal degrees of freedom but rather considered a continuum of long-wavelength fluctuations in the order parameter. With this change of perspective he made the connection between the ``transfer rates" and relaxation rates of order-parameters in viscous systems. The slow-down during a transition means large changes in the dynamic heat capacity of the system and thereby in the speed of sound. \\
Following the argument of Fixman, the slowing-down of the characteristic relaxation rate during the lipid melting transition will cause hysteresis and dissipation of heat. Even in the absence of critical phenomena, internal friction and heat conduction as introduced by Stokes \cite{Stokes1845} and Kirchhoff \cite{Kirchhoff1868}, respectively, can cause hysteresis and dissipation. However, within cooperative transitions these are secondary effects and we will disregard them for low frequencies.

\section{System Response To Adiabatic Pressure Perturbations}\label{adiabatic perturbation}
Sound is the propagation of a pressure wave that is followed by a temperature wave as a consequence of its adiabatic nature. Thermodynamically, changes in pressure ($dP$) and temperature ($dT$) couple to a change in the heat ($dQ$) of the system:
\be
dQ=\left(\pd{Q}{T}\right)_p dT+\left(\pd{Q}{P}\right)_T dp,
\label{change in heat1}
\ee 
where $c_p = (\partial Q/\partial T)_p$ is the heat capacity at constant pressure and
\begin{eqnarray}
\left( \pd{Q}{p} \right)_T & = & T \left( \pd{S}{p} \right)_T.
\label{dQdP1}
\end{eqnarray}
Due to a well-known Maxwell relation, $(\partial S/\partial p)_T=-(\partial V/\partial T)_p$, we obtain
\begin{eqnarray}
\left( \pd{Q}{p} \right)_T & = & -T\left(\pd{V}{T} \right)_p \nonumber \\
& = & -T\left(\pd{S}{T}\right)_p\left(\pd{V}{S}\right)_p \nonumber \\
& = & -c_P\left(\pd{V}{S}\right)_p.
\label{dQdP2}
\end{eqnarray}

Another Maxwell relation, $(\partial V/\partial S)_p=(\partial T/\partial p)_S$, allows us to write \eref{dQdP2} as 
\be
\left( \pd{Q}{p} \right)_T =   -c_p\left(\pd{T}{p}\right)_S.
\ee
Constant entropy implies that no heat is dissipated into the environment but only moved between different degrees of freedom within a closed system. At transitions, the Clausius-Clapeyron relation\footnote{The use of the Clausius-Clapeyron relation can be justified by the weak first-order nature of the lipid melting transition \cite{Nagle1980}.} can be used:
\be
\left( \pd{p}{T}\right)_S=\frac{\Delta H}{T \Delta V},
\label{Clausius}
\ee
where $\Delta H$ and $\Delta V$ are the enthalpy (or excess heat) and volume changes (excess volume) associated with the transition \cite{vanOsdol1989}. Note that these are constant system properties for a given transition that can be determined experimentally. 
\\\\
The change in heat (\eref{change in heat1}) can now be written as:
\be
dQ=c_p(T,p) \left( dT- \left(\frac{T \Delta V}{\Delta H}\right) dp \right)\,.
\label{change in heat}
\ee 
It is clear from \eref{change in heat} that the heat capacity acts as a transfer function that couples adiabatic changes in pressure to changes in heat. 
\\\\
\eref{change in heat} governs the equilibrium properties of the thermodynamical system. However, here we consider the propagation of sound, which is  a non-equilibrium process.  The theory of sound considers the limit of small changes in pressure and temperature for which close-to-equilibrium dynamics can be assumed.  This implies linear relations between perturbations and system responses.  For this reason it is also called `linear response theory'.
\\\\  
In any real system transfer rates are finite and changes happen in finite time. Thus, the changes in pressure and temperature can be represented as rates, and \eref{change in heat} can be rewritten as:
\be
\delta Q=\int dQ= \int c_p(t) \left[\dot{T}- \left(\frac{T \Delta V}{\Delta H}\right) \dot{p}\right] dt\,,
\label{heat2}
\ee 
where $\dot{T}=\partial T/\partial t$, and $\dot{p}=\partial p/\partial t$ are rates.
Note that $T=T_{equilibrium}$, which holds if absolute changes in temperature upon pressure changes are very small. 
\\\\
If changes in pressure or temperature happen faster than the transfer rate (or relaxation rate), the energy transferred during this change will be only  a part of the amount otherwise transferred. Considering \eref{change in heat}, the finite transfer rate will  lower the effective transfer function, in this case the heat capacity. This means that also the heat capacity must contain a relaxation term, $(1-\Psi_{c_p})$,  with $0\le\Psi_{c_p}\le1$. This function describes the equilibration of the system.  As the system approaches equilibration, $(1-\Psi_{c_p})$ approaches unity. Below, we will assume that the function $\Psi_{c_p}$ is an exponentially decaying function of time.  \eref{heat2} must then be written as a convolution: 
\be
\delta Q(t) =\int_{-\infty}^{t} \left[ c_p(\infty)+\Delta c_{p}\left(1-\Psi_{c_p}(t-t') \right) \right] \left( \dot{T}(t') - \frac{T \Delta V}{\Delta H} \dot{p} \right ) dt',
\label{heat3}
\ee 
where $\delta Q(t)$ is the change in heat, $c_p(\infty)$ is the part of the heat capacity that relaxes more rapidly than the changes in pressure and temperature considered.  In the lipid bilayer system $c_p(\infty)$ is the heat capacity contribution from lipid chains, which we consider as a background contribution. $\Delta c_{P}$ is that part of the heat capacity which relaxes on time scales of a similar order or longer than the perturbation time scale.  In the lipid membrane system this is the excess heat capacity.
In \eref{heat3} it has been assumed that the mechanisms of relaxation are the same for pressure and temperature. This assumption has been justified experimentally and numerically in the literature \cite{Ebel2001,Halstenberg1998,Heimburg1998,Grabitz2002}. 
\\\\
After partial integration of \eref{heat3}, subsequent Fourier transformation and the use of the convolution theorem, \eref{heat3}  can be transformed into (see Appendix):
\be
\delta Q = c_p(\omega)\left(T(\omega)-\frac{T \Delta V}{\Delta H}p(\omega)\right)\,. 
\label{heatfourie}
\ee
$T(\omega)$ and $p(\omega)$ can be regarded as periodic variations of temperature and pressure, respectively. We have also now introduced the frequency dependent heat capacity,
\be
c_p(\omega)= c_p(\infty)-\Delta c_{p}\int_{0}^{\infty}e^{-i\omega t}\dot{\Psi}_{c_p}(t) dt \,.
\label{cp(w)}
\ee
From \eref{cp(w)} the frequency dependent transfer function (dynamic heat capacity)\footnote{It is important note the difference between the dynamic heat capacity (frequency dependent) and the normally known equilibrium heat capacity. The equilibrium heat capacity is a constant system property whereas the dynamic heat capacity is an effective heat capacity that can be less than or equal to the equilibrium heat capacity as a consequence of the finite transfer rates in real systems.} can be found, giving a full description of how a lipid bilayer responds to adiabatic pressure perturbations.  Both $c_P(\infty)$ and $\Delta c_{p}$ are experimentally available using differential scanning calorimetry (DSC). The only unknown is the relaxation function, $\Psi_{c_p}$. 

\subsection{Relaxation Function}\label{single}
The relaxation function of the heat capacity is related to the rate of energy transfer  from the membrane to the environment. The fluctuation-dissipation theorem ensures that the rate of energy transfer is equivalent to the relaxation behavior of energy fluctuations.  Since the heat capacity is a measure of enthalpy fluctuations,  the relaxation function of the heat capacity must be the relaxation function of the enthalpy fluctuations \cite{vanOsdol1989}. 
\\\\
The relaxation behavior of the fluctuations of enthalpy in pure lipid vesicles has been considered theoretically, numerically and experimentally showing that the relaxation of enthalpy is well described by a single exponential function \cite{Grabitz2002,Seeger2007}:
\be
(H-\avg{H})(t)=(H-\avg{H})(0)\cdot \exp{\left( -\frac{t}{\tau}\right)},
\label{H(t)}
\ee
where $(H-\avg{H})(0)$ serves only as a proportionality constant and $\tau$ is the relaxation time.  For various pure lipid membranes close to melting transitions, it was further found that relaxation times are proportional to the excess heat capacity,
\be
\tau=\frac{T^2}{L}\Delta c_p \, ,
\label{tau}
\ee
where $L$ is a phenomenological coefficient. For LUV of DPPC $L=13.9\cdot10^8J \cdot K/(s\cdot mol)$ \cite{Grabitz2002}. 

\subsection{Response Function}
Using the relaxation function of the enthalpy fluctuation as the relaxation function of the dynamic heat capacity, 
\be
\Psi_{c_P}=\exp\left( -\frac{t}{\tau}\right),
\label{relax-s}
\ee
\eref{cp(w)} can be solved and the dynamic heat capacity can be determined.
\begin{eqnarray}
c_p(\omega) & = & c_p(\infty)-\Delta c_{p}\int_{0}^{\infty}e^{-i\omega t}\left(-\frac{1}{\tau}\right)e^{-\frac{t}{\tau}} dt \nonumber \\
& = & c_p(\infty) + \Delta c_{p}\left( \frac{1-i\omega \tau}{1+(\omega \tau)^2}\right) 
\label{cp(w)-final-s}
\end{eqnarray}
Note that the above derivations can be carried out with lateral pressure instead of pressure; the choice of using pressure is entirely for notational convenience. 

\section{Adiabatic Compressibility}\label{compressibility}
\begin{figure}[!h]
\centering
\includegraphics[width= 0.5 \linewidth]{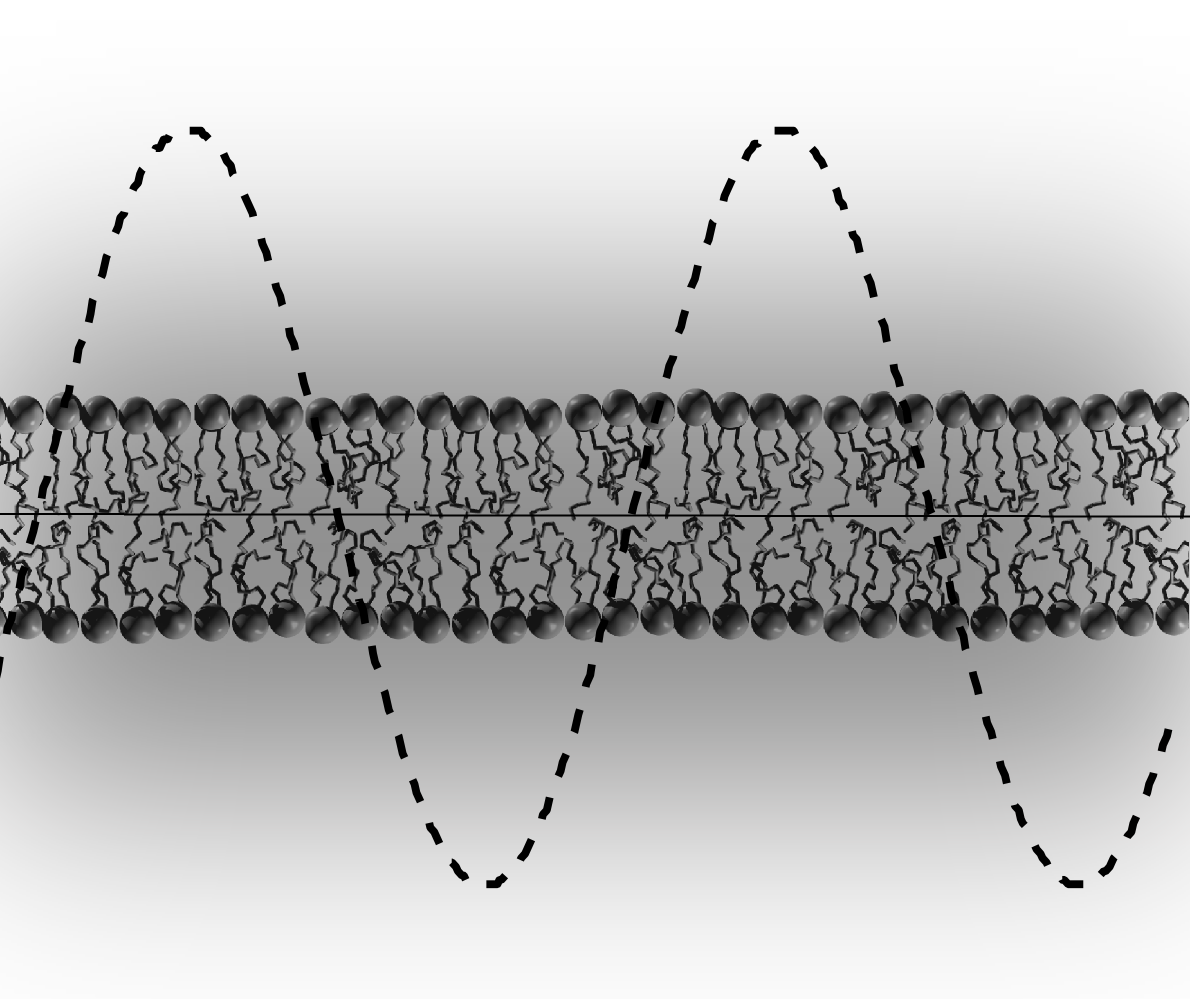}
\caption{\small \textbf{Visualization of temperature wave in the plane of a lipid bilayer. The coloring indicates heat penetrating into the surrounding water.}}
\label{t-wave}
\end{figure}

In estimating the speed of sound in the plane of a lipid membrane during the melting transition, the response of the membrane to sound (the dynamic heat capacity) must be related to the lateral adiabatic compressibility. The adiabatic lateral compressibility is defined as
\be
\kappa_{S}^{A}=-\frac{1}{A}\left(\pd{A}{\Pi}\right)_S,
\label{kappa_a0}
\ee
where $\Pi$ is the lateral pressure. The adiabatic lateral compressibility can be rewritten in the following form \cite{Wilson1957}: 
\be
\kappa_{S}^{A}=\kappa_{T}^{A}-\frac{T}{A \; c_{p}^{system}}\left(\pd{A}{T}\right)_{\Pi}^2,
\label{kappa_a}
\ee
where
\be
\kappa_{T}^{A}=-\frac{1}{A}\left(\pd{A}{\Pi}\right)_T=\kappa_{T}^A(\infty)+\frac{\gamma_{A}^2 T}{A}\Delta c_{P},
\label{kappa_T}
\ee
is the isothermal lateral compressibility, $\kappa_{T}^A(\infty)$ is that part of the isothermal lateral compressibility that relaxes 
faster than changes in the pressure and temperature considered, and $c_{p}^{system}$ is the heat capacity of the total thermodynamical system,   i.e., the lipid membrane plus the accessible surrounding aqueous medium that serves as a buffer for heat transfer. In the last equality, the empirical  proportionality $\Delta A = \gamma_A \Delta H$ has been used \cite{Ebel2001,Heimburg2005c}, with $\gamma_A=0.89$ m$^2$/J for a lipid monolayer of DPPC.
\\\\
In the literature on attenuation and dissipation of sound in critical media a different form of \eref{kappa_a} is often used to relate the dynamic heat capacity and the adiabatic compressibility, using the dynamic heat capacity as the heat capacity of the total system \cite{Bhattacharjee1997,Barmatz1968}. This can be done in a straight forward manner by employing the Pippard-Buckingham-Fairbank relations (PBFR) \cite{Pippard1956,Buckingham1961}.  The main difference between this approach and the one adopted here is that their compressible medium is three-dimensional, and the system heat capacity is that of this medium.  In contrast, the lipid membrane system is a pseudo two-dimensional (the bilayer) embedded in a three-dimensional aqueous medium that serves as a heat reservoir (see Fig. \ref{t-wave}). Therefore,  the aqueous medium contributes significantly to the features of the membrane in a frequency dependent manner.

Imagine a standing temperature wave in the bilayer. The transfer of heat from the wave to the surrounding water will be time dependent, see \fref{t-wave} for visualization. In the limit of $\omega \rightarrow 0$, the amount of water (heat reservoir) participating will effectively go to infinity. In the other extreme, ($\omega \rightarrow \infty$), no heat will be transferred to the surrounding heat reservoir.  Evidently, the heat capacity 
of the total system is frequency dependent:
\be
c_{p}^{system}(\omega)=c_p^{lipid}+c_p^{reservoir}(\omega)
\label{cp-system}
\ee 
where $c_P^{lipid}=\Delta c_P+c_P(\infty)$ is the complete heat capacity (in equilibrium) of the lipid membrane and $c_P^{reservoir}(\omega)$ is the heat capacity of the participating heat reservoir. In this approach it is the size of the contributing heat reservoir that is frequency dependent. 
\\\\
Using the proportionality relation $\Delta A = \gamma_A \Delta H$ in \eref{kappa_a} and assuming that $(\partial A/\partial T)_{\Pi}$ in the chain melting transition region is completely dominated by the transition associated change in area, the following approximation can be made \cite{Halstenberg1998}:
\begin{eqnarray}
\kappa_{S}^{A} &\approx& \kappa_{T}^A(\infty)+\frac{\gamma_{A}^2 T}{A}\Delta c_{P}-\frac{\gamma_{A}^2 T}{A}\frac{(\Delta c_{P})^2}{c_{P}^{system}} \nonumber \\
& = & \kappa_{T}^A(\infty)+\frac{\gamma_{A}^2 T}{A}\left(\Delta c_{P}-\frac{(\Delta c_{P})^2}{c_{P}^{system}}\right).
\label{kappa_a_a}
\end{eqnarray}
The parenthesis has the units of a heat capacity and is frequency dependent through the frequency dependence of the size of the associated heat reservoir. We pose as an ansatz here that this parenthesis is the effective heat capacity of the lipid membrane in a finite adiabatically isolated heat reservoir, which is equivalent to the dynamic heat capacity of the lipid membrane following the above argument. 
\be
\Delta c_P (\omega)=\Delta c_{P}-\frac{(\Delta c_{P})^2}{c_{P}^{system}}.
\label{cp-finite}
\ee
Numerical justification of this ansatz will be published at a later point.
\\\\
Using this ansatz, the dynamic heat capacity can be related directly to the adiabatic lateral compressibility through \eref{kappa_a_a}:
\be
\kappa_S^A=\kappa_T^A(\infty)+\frac{\gamma_{A}^2 T}{A}\cdot \Delta c_P(\omega),
\label{kappa}
\ee 
where the $\Delta c_P(\omega)$ is the dynamic heat capacity without background. In this equation, we use the compressibility of the lipid bilayer, which is half of that of the lipid monolayer. 
 

\section{Results -- The Speed of Sound}
\begin{figure}[!h]
\centering
\includegraphics[width= 0.9 \linewidth]{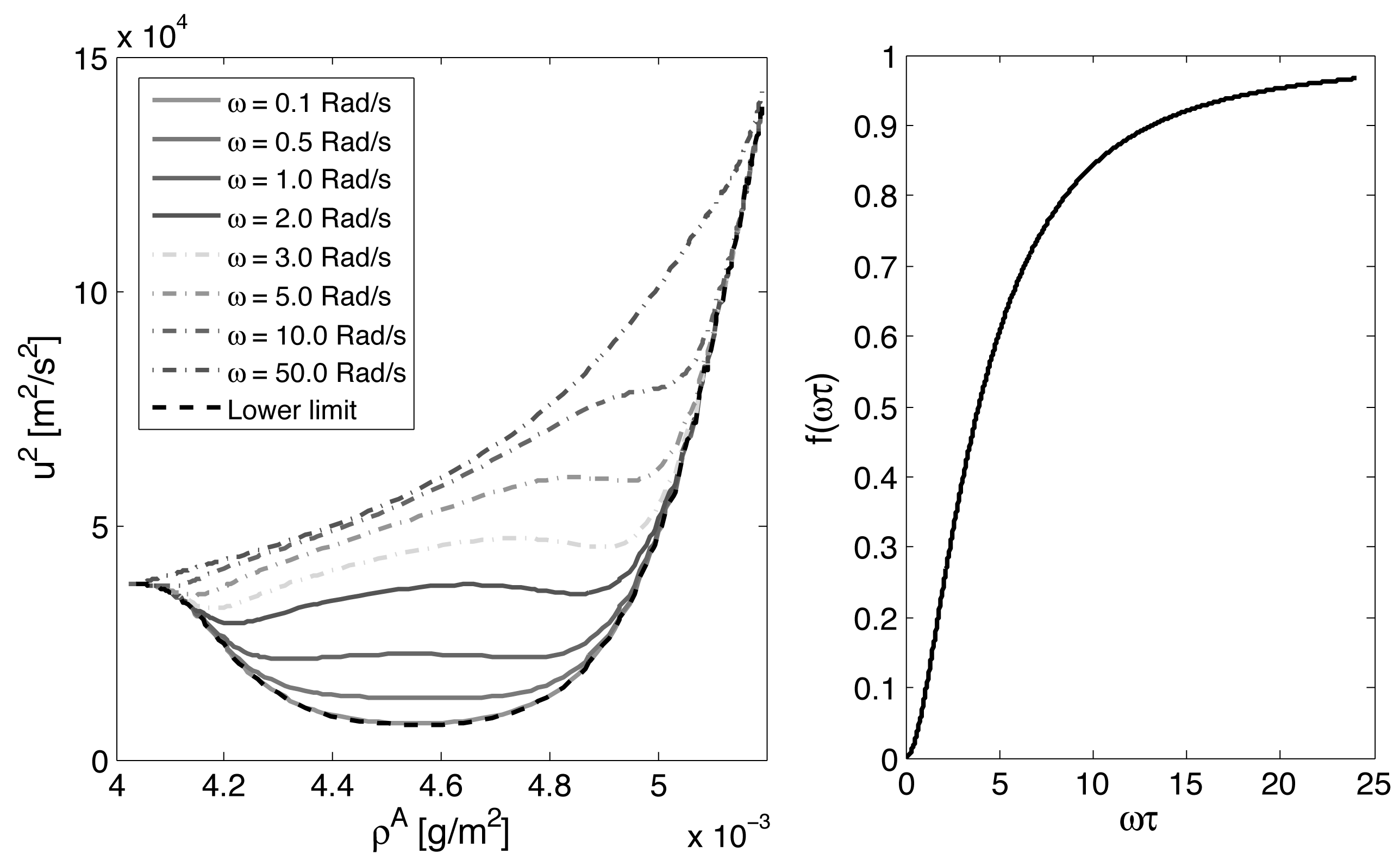}
\caption{\small \textbf{\textit{Left:} The effective lateral speed of sound squared as a function of density at different angular frequencies along with the limiting case: $\boldsymbol{\omega \rightarrow 0}$. \textit{Right:} The generic function, $\boldsymbol{f(\omega\tau)}$, that takes the effective lateral speed of sound squared, at a given lateral density, from the low frequency limit ($\boldsymbol{f(\omega\tau \rightarrow 0)=0}$) to the high frequency limit ($\boldsymbol{f(\omega\tau \rightarrow \infty)=1}$).}}
\label{velocity-s}
\end{figure}

The goal is to estimate the speed of sound and its frequency dependence in the plane of a lipid membrane. From the estimated dynamic heat capacity \eref{cp(w)-final-s}, the adiabatic lateral compressibility can found using the proposed relation (\eref{kappa}).  The lateral speed of sound can then be estimated using \eref{speed0} as:
$$
c^A=\frac{1}{\sqrt{\kappa_S^A \rho^A}} \, , \nonumber
$$
where $\kappa_S^A$ is a function of the frequency, $\omega$. The effective speed of sound is given by (\eref{real-speed}):
\be
u = \left(\frac{Re(c^A)}{| c^A |^2}\right)^{-1}. \nonumber
\ee
Using the previous two equations, one can show that
\be
u^2(\omega) =(\rho^A)^{-1} \frac{2}{Re(\kappa^A_S)+ | \kappa^A_S|}.
\label{real-speed2}
\ee
Inserting the estimated adiabatic lateral compressibility from \eref{kappa} and \eref{cp(w)-final-s}) into \eref{real-speed2}, the effective speed of sound squared takes the analytic form:
\be
u^2(\omega) = \frac{2}{\frac{1}{c_1^2}+\frac{1}{c_2^2}\frac{1}{(1+(\omega \tau)^2)}+ \sqrt{\left(\frac{1}{c_1^2}+\frac{1}{c_2^2}\frac{1}{(1+(\omega \tau)^2)} \right)^2+ \left(\frac{1}{c_2^2}\frac{\omega \tau}{(1+(\omega \tau)^2)} \right)^2}} \, ,
\label{real-speed-final}
\ee
with the notation 
\be
c_1^2 \equiv \left(\rho^A \kappa_T^A(\infty) \right)^{-1}
\label{c1}
\ee
and
\be
c_2^2(\omega) \equiv \left(\rho^A \frac{\gamma_{A}^2 T}{\avg{A}}\Delta c_{p}(\omega)  \right)^{-1} \, .
\label{c2}
\ee
Here,  $c_1$ is the lateral speed of sound of the membrane outside of the transition, and $c_2$ is the component of the lateral speed of sound related to the lipid melting transition.
\\\\
All variables in \eref{real-speed-final} can be found from the excess heat capacity of the lipid melting transition and the fluid fraction,\footnote{The fluid fraction is the fraction of a considered lipid system that is in the fluid phase.} which can be obtained using differential scanning calorimetry (DSC). The area, the lateral density and the background isothermal compressibility are all directly related to the fluid fraction \cite{Heimburg1998}. The relaxation time can be estimated from its phenomenological proportionality relation to the excess heat capacity, \eref{tau}. The proposed analytic expression for the effective speed of sound (\eref{real-speed-final}) is shown in \fref{velocity-s}, where the excess heat capacity and the fluid fraction are taken from Monte Carlo simulations of the lipid melting transition in LUV of DPPC. The simulation have been carried out in a manner similar to that described in \cite{Heimburg1996a}.

The frequency dependence of the speed of sound is described by the function, $f(\omega \tau)$ with $0\le f(\omega\tau)\le 1$, defined by
\be
u^2(\omega\tau)=u_0^2 + (u_{\infty}^2 - u_0^2) \, f(\omega\tau)
\label{f_w}
\ee 
where $u_0 = u(\omega\tau \to 0)$ and $u_\infty \equiv u(\omega\tau \to \infty)$.  From \eref{real-speed-final} we see that
\be
u_0^2 = \left(\frac{1}{c_1^2}+\frac{1}{c_2^2}\right)^{-1}\nonumber
\ee
\be
u_{\infty}^2 = c_1^2 \, .
\label{f_w_2}
\ee
See \fref{velocity-s} (\textit{right}).  The generic function $f$ was chosen to be a function of the dimensional less quantity 
$\omega \tau$ rather than $\omega$ in order to render it independent of the lateral density. 

\subsection{Dispersion Relation}
\begin{figure}[!h]
\centering
\includegraphics[width= 0.7 \linewidth]{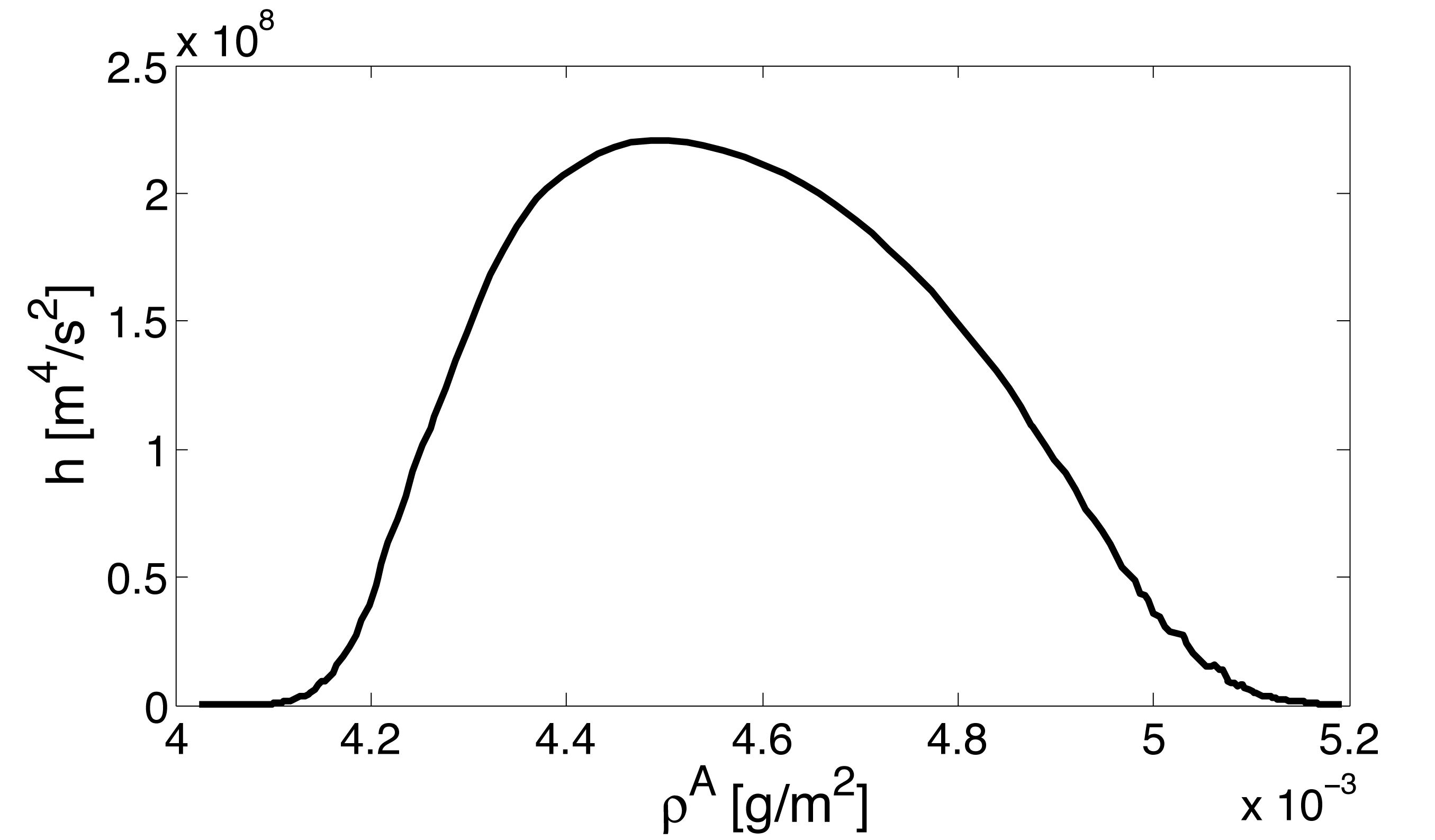}
\caption{\small \textbf{The dispersion parameter, $\boldsymbol{h}$, as a function of lateral density for LUV of DPPC, based on the proposed expression for the lateral speed of sound.}}
\label{h-fig}
\end{figure}

In the Soliton model described by \eref{sound3}, dispersion was assumed to be small and independent of the lateral density due to the lack of detailed information of the frequency dependence of the speed of sound as a function of density. Using the considerations of the previous paragraphs, we can now estimate the dispersion in lipid membranes. In the Soliton model the extent of dispersion is described by the parameter, $h$.  Assuming that dispersion is small, $h$ can be related to the lateral speed of sound as 
\be
u^2 \approx c_0^2+\frac{h \omega^2}{c_0^2} + ... \, ,
\label{dis}
\ee 
where $c_0^2\equiv u^2(\omega \rightarrow 0)=(1/c_1^2+1/c_2^2)^{-1}$. \eref{dis} corresponds to a Taylor expansion of the lateral speed of sound squared to second order.\footnote{The first order term is zero since the speed of sound squared is symmetric around $\omega = 0$.}. Expanding \eref{real-speed-final} to second order,
\be
u^2 \approx c_0^2+c_0^4\frac{3c_1^2+4c_2^2}{4c_2^2(c_1^2+c_2^2)}\omega^2\tau^2,
\label{taylor-u}
\ee
we see that the dispersion parameter has the form:
\be
h = c_0^6\frac{3c_1^2+4c_2^2}{4c_2^2(c_1^2+c_2^2)}\tau^2 \, .
\label{h}
\ee 
Using the excess heat capacity and the fluid fraction for large unilamellar vesicles of DPPC as used in \fref{velocity-s}, we can estimate the density dependence of the dispersion parameter $h( \rho_A )$  as shown in \fref{h-fig}. 

Here, the density of the fluid phase is approximately $4\cdot 10^{-3}$ g/m$^2$, the maximum of the dispersion parameter corresponds to the chain melting transition maximum, and the density of the gel phase is $5\cdot 10^{-3}$ g/m$^2$.  It is clear that the dispersion parameter is strongly dependent on the lateral density of the membrane. \\
The density dependent dispersion parameter, $h(\rho^A)$, will finally enter the differential equation (\eref{sound3}) for the propagating nerve pulse. In the original treatment, $h$ was considered an adjustable constant that determined the time scale of a solitary pulse in nerve axons. In the present extension, $h(\rho^A)$ is fully determined by the cooperative nature of the membrane system and does not contain adjustable parameters. Preliminary calculations indicate that this dispersion parameter will yield a natural time scale for the propagating soliton in nerve axons.

\section{Discussion}

The response of lipid membranes to adiabatic periodic pressure perturbations (sound) is closely related to the relaxation behavior of the system \cite{Herzfeld1928, Fixman1962}. Using thermodynamics and linear response theory, we have described the response of the lipid membrane to a perturbation with the assumption that the relaxation function has a simple exponential dependence on time. We obtain a form for the dynamic heat capacity which can be understood as the effective heat capacity when the lipid membrane is subject to periodic adiabatic pressure perturbations. The dynamic heat capacity was then related to the adiabatic lateral compressibility using the idea that the size of the associated water reservoir is frequency dependent \cite{Halstenberg1998}.  The adiabatic lateral compressibility was then used to obtain an expression for the effective speed of sound as a function of frequency.

\begin{figure}[!h]
\centering
\includegraphics[width= 0.9 \linewidth]{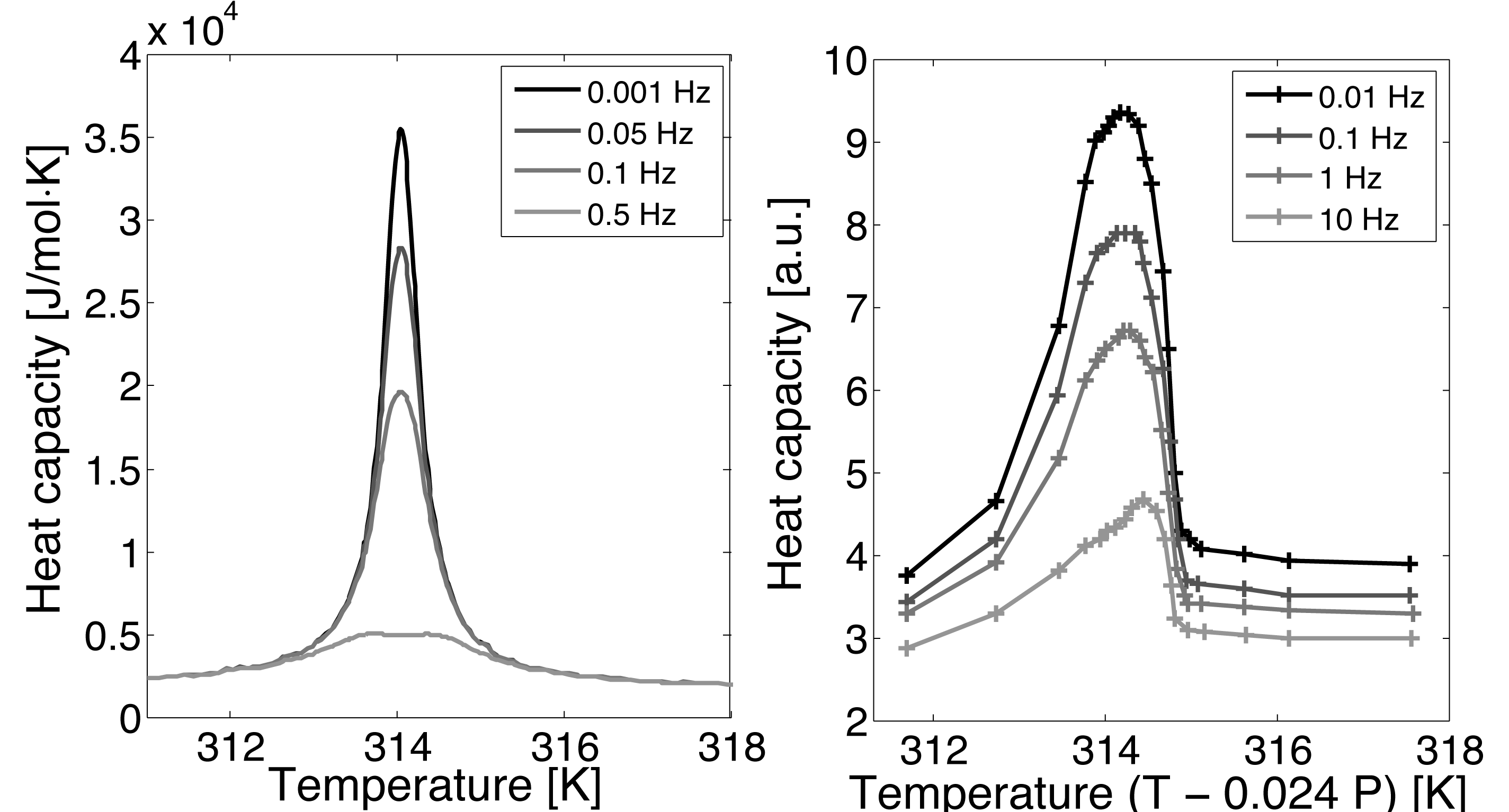}
\caption{\small \textbf{\textit{Left:} The calculated dynamic heat capacity for LUV of DPPC at different frequencies. \textit{Right:} The effective heat capacity profiles for LUV of DPPC at different frequencies, measured by Van Osdol \emph{et al.} \cite{vanOsdol1991a}. The measured effective heat capacities have not been corrected for contributions from the experimental setup, and a direct comparison is therefore not possible. The theoretical dynamic heat capacity shows the same qualitative features as the measurements --- a dramatic decrease in the height of the excess heat capacity with increasing frequency and a relatively constant width. The difference in frequency scales seen in the two panels is due to an estimated difference of more than a factor of 10 in the characteristic relaxation time. Frequencies are given in units of  Hz = (1/2$\pi$) rad/s .}}
\label{vanosdol}
\end{figure}

The major assumption in our approach concerns the nature of the relaxation function.  We have previously studied the relaxation behavior of the lipid membrane in the vicinity of the melting transition at low frequencies.\footnote{The time resolution of experiments from our group is $0.3\,s \sim 3.3$\,Hz. Relaxation profiles on longer time scales are well approximated by a single exponential decay\cite{Grabitz2002, Seeger2007}. }This means that the lipid melting transition is assume to be \emph{non}-critical. The single exponential relaxation behavior should, however, only be considered as a low frequency approximation. In a number of ultrasonic experiments it has been shown that a single exponential is insufficient to describe the dynamics of the cooperative processes involved in lipid melting in the ultrasonic regime \cite{Mitaku1982,Halstenberg1998,Schrader2002,Halstenberg2003}. In these ultrasonic experiments, some phase transition phenomena are even apparent in the MHz regime.  Single exponential relaxation behavior, and thereby the validity of the estimated speed of sound, is thus limited to frequencies comparable to the relaxation rate or lower. 
\\\\
Van Osdol \emph{et al.} \cite{vanOsdol1991a} have made adiabatic pressure perturbation experiments on unilaminar and multilaminar vesicles of DPPC.  They studied relaxation behavior of the lipid membrane by measuring the frequency dependence of the effective heat capacity and the compressibility as a function of frequency. Although the data available for unilamellar vesicles is very limited and has large errors, it can still serve to illustrate qualitative tendencies of the effective heat capacity, see \fref{vanosdol}, that are similar to the theoretical results reported here.  
The effective frequency dependence of the speed of sound shown in \fref{velocity-s} is dominated by the cooperative properties of the lipid melting transition of DPPC. In this model system the relaxation time during the transition is as slow as seconds.  In biological membranes such as membranes of nerves, realistic characteristic relaxation times can be assumed to be of the order of $1-100$ ms. This change in relaxation times between the model system and biological membrane expands the upper limit of the frequency range for which our approach is likely to be valid from several Hz to several kHz regime, assuming that the general behavior of pure lipid and biological membranes is otherwise similar.  Since the duration of a nerve pulse is roughly 1 ms, the relevant frequency components contained in a nerve pulse can be estimated to be 1 kHz or less.  The relevant frequency range for nerve pulses is thus covered by our proposed expression for the effective speed of sound.  The present results may thus provide useful insights regarding sound propagation in an otherwise inaccessible regime and can extend our understanding of the nature of nerve signals.

In future studies, the linear response theory described in this chapter will help to define an intrinsic length scale of the electromechanical soliton proposed by us as an alternative description for the nervous impulse.


\begin{appendix}

\section{Derivation of the dynamic heat capacity using the convolution theorem}
The purpose of this appendix is to provide additional details in the derivation of the frequency dependent heat capacity given 
in \eref{cp(w)-final-s} starting from \eref{heat3}.  The change in heat is a convolution of the applied perturbation and the relaxation of the transfer function --- the effective heat capacity. The perturbation is well defined at all times and can safely be assumed to be zero for $t \to - \infty$.  The relaxation function is only defined from $ [0, \infty]$, where $ t=0 $ is the time at which the system starts to equilibrate. The relaxation function, $\Psi$, is chosen such that $\Psi(t \to 0)=1$ and $\Psi(t \to \infty)=0$.  To accommodate the chosen form of the relaxation function the convolution can be written as following:

\begin{eqnarray}
\delta Q(t) & = & \int_{-\infty}^{t}(c_p(\infty)+\Delta c_p(1-\Psi(t-t'))) \left( \dot{T}(t')-\frac{T_0 \Delta V}{\Delta H} \dot{p}(t') \right)dt' \label{dq0}\\
& = & \int_{-\infty}^{t} g(t-t')\dot{f}(t')dt' ,
\label{dq}
\end{eqnarray}
where $g(t-t')$ is the transfer function and $\dot{f}(t')$ is the perturbation. Note that $\dot{f}(t)=df(t)/dt$, $c_p(\infty)$ is the component of the heat capacity not associated with the melting transition, and $T_0$ is the equilibrium temperature. \\\\
Integration by parts allows us rewrite \eref{dq} to the following form:
\be
\delta Q(t)=\left[ g(t')\int \dot{f}(t')dt' \right]^t_{-\infty}-\int_{-\infty}^{t}\left( \int \dot{f}(t'')dt'' \right) \dot{g}(t-t')dt' \, .
\label{dq2}
\ee
The first term in \eref{dq2} takes the form:
\begin{eqnarray}
\left[ g(t')\int \dot{f}(t')dt' \right]^t_{-\infty} & = & \left[ g(t') f(t') \right]^t_{-\infty} \nonumber \\
&=& g(t)f(t)-g(-\infty)f(-\infty) 
\label{fq1}
\end{eqnarray}
where 
\be
f(t')=(T(t')-T_0)-\frac{T_0 \Delta V}{\Delta H} (p(t')-p_0) \, .  \nonumber
\ee 
Assuming that the system is in equilibrium as $t \to -\infty$ and $f(t' \to  -\infty)=0$, simplifies \eref{fq1}.
\be
g(t)f(t)-g(-\infty)f(-\infty)=c_p(\infty) f(t) .
\ee
The second term in \eref{dq2} can be rewritten by changing the variable to $t''=t-t'$.
\be
\int_{-\infty}^{t} \left( \int \dot{f}(t')dt' \right)\dot{g}(t-t')dt'=-\int_{0}^{\infty} f(t-t'')\dot{g}(t'')dt'' \, ,
\ee
where the integration limits have been changed accordingly.\\\\ 
Since we are interested in sinusoidal perturbations, we consider the Fourier transform of \eref{dq0} and find:
\begin{eqnarray}
\delta \hat{Q}(\omega) & = & \int^{\infty}_{-\infty}\delta Q(t) e^{-i \omega t} dt \label{dq1} \\
& = & \int^{\infty}_{-\infty}  \left( c_p(\infty)f(t)+\int_{0}^{\infty}f(t-t'')\dot{g}(t'')dt''\right) e^{-i \omega t}  dt .
\label{dq3}
\end{eqnarray}
The Fourier transform of the first term in \eref{dq3} can be carried out without complications:
\be
c_p(\infty)\int^{\infty}_{-\infty} f(t) e^{-i \omega t} dt=c_p(\infty) \hat{f}(\omega) .
\label{dq5}
\ee
The second term of \eref{dq3} can be rewritten as follows:
\be
\int^{\infty}_{-\infty} \int_{0}^{\infty}f(t-t'')\dot{g}(t'') e^{-i \omega t} dt'' dt=\int_{0}^{\infty} \dot{g}(t'') \int^{\infty}_{-\infty}  f(t-t'') e^{-i \omega t}  dt  dt'' 
\ee
Changing variables again, $t'=t-t''$, the Fourier transform of the second term in \eref{dq3} can be split into two terms:
\begin{eqnarray}
\int_{0}^{\infty} \dot{g}(t'') \int^{\infty}_{-\infty}  f(t-t'') e^{-i \omega t}  dt'' dt &= &\int_{0}^{\infty} \dot{g}(t'') \int^{\infty}_{-\infty}  f(t') e^{-i \omega (t'+t'')}  dt'  dt'' \nonumber \\
&=& \int_{0}^{\infty} \dot{g}(t'') e^{-i \omega t''} dt'' \int^{\infty}_{-\infty}  f(t') e^{-i \omega t'}  dt' \nonumber \\
&=& \hat{f}(\omega) \int_{0}^{\infty}  e^{-i \omega t} \dot{g}(t) dt \, .
\label{dq4}
\end{eqnarray}
This is known as the Convolution theorem. From \eref{dq4} and \eref{dq5}, \eref{dq1} can be written as:
\be
\delta \hat{Q}(\omega)=\left( c_p(\infty)+\int_{0}^{\infty}  e^{-i \omega t} \dot{g}(t) dt )\right) \hat{f}(\omega) ,
\ee
where 
\be
\hat{f}(\omega)=\hat{T}(\omega)-\frac{T_0 \Delta V}{\Delta H} \hat{p}(\omega) \ \  
{\rm and}\ \ \dot{g}(t)=-\Delta c_p \dot{\Psi}(t) \, . \nonumber
\ee
The Fourier transform of \eref{dq} takes the final form:
\begin{eqnarray}
\delta \hat{Q}(\omega)&=&\left( c_p(\infty)-\Delta c_p \int_{0}^{\infty}  e^{-i \omega t} \dot{\Psi}(t) dt )\right) \left(\hat{T}(\omega)-\frac{T_0 \Delta V}{\Delta H} \hat{p}(\omega)\right) \\
&=& c_p(\omega) \left(\hat{T}(\omega)-\frac{T_0 \Delta V}{\Delta H} \hat{p}(\omega)\right) .
\end{eqnarray}
Using $\Psi(t)=\exp(-t/\tau)$, the dynamic heat capacity, $c_p(\omega)$, is found to be 
\begin{eqnarray}
c_p(\omega) &=& c_p(\infty)+\frac{\Delta c_p}{\tau} \int_{0}^{\infty}  e^{-i \omega t} e^{-t/\tau}  dt \\
&=& c_p(\infty)+\Delta c_p \left( \frac{1-i\omega\tau}{1+(\omega\tau)^2} \right) \, ,
\label{dq6}
\end{eqnarray}
which has the form of a Debye relaxation term.

\end{appendix}


\end{document}